# A pseudo-capacitive chalcogenide-based electrode with dense 1-dimensional nanoarrays for enhanced energy density in asymmetric supercapacitors

Young-Woo Lee[a,1], Byung-Sung Kim[a,1], Jong Hong[a], Juwon Lee[a], Sangyeon Pak[a],

Hyeon-Sik Jang[b], Dongmok Whang[b], SeungNam Cha[a,*], Jung Inn Sohn[a,*] and Jong Min Kim[a,c]

[a] Department of Engineering Science, University of Oxford, Oxford OX1 3PJ, United Kingdom.

[b] School of Advanced Materials Science and Engineering, SKKU Advanced Institute of
  Nanotechnology (SAINT), Sungkyunkwan University, Gyeonggi-Do 16419, Republic of Korea.

[c] Electrical Engineering Division, Department of Engineering, University of Cambridge, Cambridge
  CB3 0FA, United Kingdom.

[1] These authors contributed equally to this work.

* Corresponding author. . Tel: +44-1865-273912. Fax: +44-1865-273010.

  E-mail address: seungnam.cha@eng.ox.ac.uk; junginn.sohn@eng.ox.ac.uk.

# ABSTRACT

To achieve the further development of supercapacitors (SCs), which have intensively received attention as a next-generation energy storage system, the rational design of active electrode materials with electrochemically more favorable structure is one of the most important factors to improve the SC performance with high specific energy and power density. We propose and successfully grow copper sulfide (CuS) nanowires (NWs) as a chalcogenide-based electrode material directly on a Cu mesh current collector using the combination of a facile liquid-solid chemical oxidation process and an anion exchange reaction. We found that the as-prepared CuS NWs have well-arrayed structures with nanosized crystal grains, a high aspect ratio and density, as well as a good mechanical and electrical contact to the Cu mesh. The obtained CuS NW based electrodes, with additional binder- and conductive material-free, exhibit a much higher areal capacitance of 378.0 mF $cm^{-2}$ and excellent cyclability of an approximately 90.2% retention during 2000 charge/discharge cycles due to their unique structural, electrical, and electrochemical properties. Furthermore, for practical SC applications, an asymmetric supercapacitor is fabricated using active carbon as an anode and CuS NWs as a cathode, and exhibits the good capacitance retention of 91% during 2000 charge/discharge processes and the excellent volumetric energy density of 1.11 mW h $cm^{-3}$ compared to other reported pseudo-capacitive SCs.

**Keywords:** Copper chalcogenide, Nanowire, Nano-sized grain, Pseudocapacitance, Asymmetric supercapacitor

**INTRODUCTION**

Supercapacitors (SCs), such as pseudocapacitor and electric double-layer capacitors (EDLCs) that can bridge the potential gap between conventional rechargeable batteries and transitional electrostatic capacitors, have been considered as a promising building block for next generation energy storage systems that require the high specific energy and power density for wide-ranging energy/power supplies, electric vehicles, and portable devices.[1-3] Pseudocapacitors, eminently, can achieve a high specific capacitance compared to EDLCs because they are basically operated by electrochemical Faradaic redox reactions, leading to much greater charge storage through the chemical intercalation/deintercalation process of anion/cation ions into electrodes.[4,5] Accordingly, high capacitive electrodes for pseudocapacitors, coupled with both high energy and power densities, have been desired and investigated by focusing primarily on transition metal based oxide/hydroxide materials (*i.e.,* $MO_x$, $M(OH)_x$, and $M_1M_2O_x$; M=Co, Ni, Cu, W, Mo, etc.)[6-10] owing to their richer redox chemical valences as well as on their architecture engineering[11-13] to enhance high surface area/electrolyte contact areas and to induce fast charge transfer rates.

Copper oxide/hydroxide-based materials are one of the attractive pseudo-electrodes to enhance a charge storage ability in the pseudocapacitors due to their two-electron redox reactions.[14,15] In addition, in order to reach a high areal capacitance and good cyclability of pseudo-electrodes, several strategies have been developed for the production of architecture-controlled, nano-sized, and complex materials to induce more intercalation active sites, effortless charge transfer pathway, and improved electrical conductivity.[16-18] Despite many efforts for the high electrochemical performance of the oxide/hydroxide-based pseudo-electrodes, there still remain the challenges, *i.e.*, low areal capacitance, poor rate performance, unacceptable energy density, and unstable cyclability because of their low intrinsic electrical conductivity and low structural stability during pseudo-charge/discharge processes.[19,20]

In this regard, copper chalcogenide based materials have been recently spotlighted, due to their high intrinsic electrical conductivity and fast electron/charge transfer rates as well as two electron redox reactions. Recently, Lei et al. demonstrated that copper chalcogenide/polypyrrole prepared by an *in-situ* oxidative polymerization approach exhibits a superior specific capacity of 171.2 mF cm$^{-2}$ and excellent cycling stability.[21] Also, Huang et al. reported that the copper chalcogenide nanosheets synthesized using an one-step solvothermal process show the enhanced areal capacitance.[22] However, fundamental researches to enhance pseudo-capacitive performances with copper chalcogenide based materials are still in the beginning stage and need much attention for achieving the high areal capacitance and energy density in their practical SC applications.

Herein, we present a facile synthesis approach to develop the dense arrays of copper sulphide (CuS) nanowires (NWs) with nanosized crystal grains with high grain boundaries (GBs), as attractive copper chalcogenides materials for SCs, directly on the Cu mesh current collector and investigate the interrelations between their structural and electrochemical properties in pseudocapacitors. We also propose that superior electrochemical pseudo-capacitive behavior of the well-defined CuS NWs based electrodes could be attributed to the synergistic effects that might be explained with the following major viewpoints: 1) the directly grown CuS NWs on the Cu mesh, without the need of conducting additives and binder materials, can provide stable adhesion, low contact resistance, and the fast charge transfer channels, thereby can induce the high areal capacitance and volumetric energy density as well as acceptable stability; 2) the CuS nanostructures consisted by nanosized grains with high GBs can induce large contact area for electrolytes, thus facilitating the opened diffusion path favorably for OH$^-$ ions and providing the fast charge transfer rates; 3) the CuS phase with the excellent electrical conductivity and good crystallinity can induce superior electrochemical cyclability and high structural stability. In addition, we design the asymmetric supercapacitor (ASC) with active carbon (AC) as the anodic material and CuS NWs as the cathodic material to demonstrate comparable volumetric energy density of 1.11 mAh cm$^{-3}$ and good cyclability of 91% during 2000 charge/discharge processes to those of commercial SCs and other reported pseudo-capacitive SCs.

**RESULTS AND DISCUSSION**

To obtain highly uniform CuS NWs directly grown on a Cu mesh, we used a sequential two-step synthetic scheme involving (1) the solution growth of Cu(OH)$_2$ and (2) the sulfurization process for CuS, which are schematically illustrated in Fig. 1a. First, Cu(OH)$_2$ NWs were directly grown on the Cu mesh using a liquid-solid chemical oxidation process at room temperature for 60 min. Here, it should be noted that, the piece of an immersed Cu mesh not only provides Cu sources for the formation of Cu(OH)$_2$ NWs but also acts as the current collector, which allows to achieve improved adhesion, structural stability and contact resistance at the interface between the electrode material and the current collector. The obtained Cu(OH)$_2$ NWs presented a high crystallinity (Fig. 1b) and all X-ray diffraction (XRD) patterns were well indexed to the crystal phase of Cu(OH)$_2$ (JCPDS No. 13-0420). Subsequently, CuS NWs were then fabricated via the dip-coating of as-prepared Cu(OH)$_2$ NWs surface with thiourea as sulfur sources, followed by the sulfurization process. As a result, Cu(OH)$_2$ phases were perfectly converted to CuS phases as shown by XRD patterns (JCPDS No. 06-0464). Furthermore, to investigate the morphology of CuS NWs, we carried out a field emission scanning electron microscope (FE-SEM) analysis. Figs. 1c and 1d clearly show that dense CuS NWs were well arrayed and separated from each 1-D nanostructure with a high aspect ratio on the Cu mesh. Moreover, there was no observation of any structural agglomeration and collapse resulting from the sulfurization process involving sulfur anion exchange reactions, as exhibited in Fig. 1d.

To further identify the detailed crystal structure and elemental distribution of CuS NWs, we performed transmission electron microscopy (TEM) examinations. As shown in Fig. 2a, the CuS NW exhibited a well-defined 1-dimensional (D) nanostructure with the homogeneous elemental distributions of Cu and S atoms. In particular, the high-resolution TEM image reveals that the CuS NW is composed of polycrystalline small grains with the size of approximately 5~6 nm, providing the high grain boundaries, and the interplanar spacing of single crystal grains is ~ 0.28 nm corresponding to {103} facets in the hexagonal crystal structure of CuS, in Fig. 2b. This finding suggests that the well-arrayed 1-D nanostructures with nanosized grains and high GBs can provide the fast charge transfer channels and can also facilitate the opened diffusion path of OH$^-$ ions for superior electrochemical reactions. In addition to the crystal and geometrical structure, since the chemical state of active materials would be an important key factor playing a crucial role in affecting electrochemically pseudo capacitive behavior, the X-ray photoelectron spectroscopy (XPS) analysis was performed to determine the surface chemical state in CuS NWs. We clearly observed Cu $2p_{3/2}$ peaks located at 932.8 and 933.8 eV being assigned to the metallic Cu and Cu-S state, respectively, revealing the presence of $Cu^{2+}$ states, as show in Fig. 2d. Moreover, we found that the peak of S $2p_{3/2}$ with the binding energy at 162.0 eV corresponds to the Cu-S bond, further supporting the evidence for the $Cu^{2+}$ state (Fig. 2e). A comparison of structural and chemical analyses results implies that directly grown CuS NWs on the Cu mesh can exhibit the enhanced pseudo-capacitive behavior during the charge/discharge process for the excellent electrochemical performance due to their highly dense array of the 1-D nanostructures with nanosized grains and high electrical conductivity as well as $Cu^{2+}$ states, allowing for superior electrochemical reactions by providing favorable pseudo-active sites.

The pseudo-behavior properties of CuS NWs were investigated through cycling voltammetry (CV) and galvanostatic charge/discharge analyses using a three-electrode system with CuS NWs on the Cu mesh directly used as a working electrode. Fig. 3a presents CV curves obtained from CuS NWs in different upper potential windows from 0.2 to 0.5 V at a scan rate of 5 mV s$^{-1}$. It is shown that the shape of all CV curves of CuS NWs differs distinctly from the typical rectangular shape of double layer capacitances, indicating that the electrochemical capacitive behavior is mainly governed by the redox reactions. Explicitly, when the upper potential limit was increased from 0.2 to 0.5 V, we observed a pair of the redox peaks, originating from the reversible Faradic redox reaction.[14,15,23,24] Additionally, good symmetric curves indicate the excellent redox-reversibility of CuS NWs. Fig. 3b shows comparative CV results of CuS NWs and $Cu(OH)_2$ NWs. The enclosed area of CuS NWs is approximately 2.3 times bigger than that of the $Cu(OH)_2$ NWs, indicating that the CuS NWs exhibit the largely enhanced pseudo-capacitance after the sulfurization of as-grown $Cu(OH)_2$ NWs. To further demonstrate the superior electrochemical capacitive performance of CuS NWs based electrodes, the areal capacitance was calculated from galvanostatic charge/discharge curves measured in a potential window of 0 to 0.5 V at different current densities ranging from 2 to 20 mA $cm^{-2}$. At a discharging current density of 2 mA $cm^{-2}$ (Fig. 3c), the areal capacitance of CuS NWs (378.0 mF $cm^{-2}$) is 2.2 times higher than that of $Cu(OH)_2$ NWs (172.4 mF $cm^{-2}$), which is in good agreement with CV results. Furthermore, even at a high charge/discharge rate of 20 mA $cm^{-2}$, the CuS NWs exhibited a superior capacitance retention of 71.2 % as shown in Fig. 3d. Here, note that to the best of our knowledge, these areal capacity and retention performance are excellent values in comparison with those of $Cu(OH)_2$ NWs and other reported Cu-based pseudo-capacitive electrodes,[21,22,24] as summarized in Fig. 3d. These results are attributed to the structural synergistic effects, as schematically illustrated in Fig. 3e: (1) the direct growth of active materials from core metal sources can minimize interfacial resistant and facilitate charge transfer at the interface between the Cu current collector and CuS NWs; (2) the dense CuS NWs with well-ordered 1-D nanostructures and the relatively high conductive CuS phase structure can induce the fast charge/electron transfer rate via the efficient electron channel; (3) the nanosized crystal grains with high GBs distributed along NWs can provide the large contact area of electrolyte and the opened diffusion path of $OH^-$ ions.

The cyclic performance is also an important factor to evaluate the ability of a SC. Fig. 3f shows that the CuS NWs based electrode exhibited the superior cyclic performance up to 2000 charge/discharge cycles at a current density of 10 mA cm$^{-2}$. The reversible capacitance retention (90.2% of the initial maximum capacitance) of CuS NWs is substantially much better than that of previously reported Cu-based composites, such as the CuS@PPy composite (88% retention up to 1000 cycles at 1 A g$^{-1}$),[21] the CuS nanosheets (75.4% retention up to 500 cycles at 1 A g$^{-1}$),[22] the CuS embedded within porous octahedral carbon (approximately 80% retention up to 2000 cycles at 5 mV s$^{-1}$),[25] and the CuO nanoflowers (84% retention up to 2000 cycles at 100 mV s$^{-1}$).[26] To clearly understand the electrochemical and structural cycling stability of our CuS NWs, electrochemical impedance spectroscopy (EIS) was used and Nyquist plots were obtained in the frequency range of 10 mHz to 100 kHz before and after 2000 cycling tests in Fig. 3g. Before the cycling test, the charge transfer resistance ($R_{ct}$) of CuS NWs was determined to be 0.11 Ω, which may be attributed to the low contact resistance and fast charge transfer rate resulting from directly grown CuS NWs on the Cu mesh and the well-arrayed 1-D nanostructures with nanosized grains. Furthermore, after a 2000 cycling test, the CuS exhibited a slight increase in the $R_{ct}$, but still maintained a very low value. The structural stability of CuS NWs was further confirmed by Raman analyses before and after the cycling test, as shown in Fig. 3h. Initially, as expected, the two characteristic Lorentzian peaks were observed around ~260.4 and ~468.8 cm$^{-1}$ corresponding to two different vibrational (stretching) modes related to the covalent S-S bond and Cu-S bond, respectively. After a cycling test, noticeably the Raman spectrum of CuS NWs exhibited that the two dominant peaks related to a CuS phase remain unchanged, confirming that the CuS NWs have a superior structural stability during the charge/discharge process. However, additional peaks with the relatively very weak intensity were also observed near ~ 288.0, 335.2, and 617.7 cm$^{-1}$ related to a CuO phase, which might be partially formed via the intercalation process of OH$^-$ ions during the galvanostatic charge/discharge process. Thus, we believe that the remarkably superior pseudo-behavior properties of the electrode, *i.e.,* high areal capacitance and excellent cycling performance, are mainly attributed to the unique features of the CuS NWs with superior electrical conductivity, favorable structural architecture and phase, and electrochemical stability.

To further investigate practical possibility for SC applications on the basis of previous electrochemical results and discussion, we fabricated an ACS using the AC as an anodic material and the CuS NWs as a cathodic material (inset in Fig. 4e) by considering the charge balance between two electrodes to reach the ideal energy storage performance of an ACS.[27] For the comparison and estimation of operating potential ranges, CV curves of the AC and the CuS NWs were first characterized (Fig. 4a). From the sum of the potential ranges of these two electrodes, the potential window of the ACS was estimated to be 1.5 V. Fig. 4b presents CV curves of the AC//CuS ASC in different upper potential windows from 0.6 to 1.5 V at a scan rate of 50 mV s$^{-1}$, indicating that this ACS system can be operated up to 1.5 V under stable electrochemical behavior. The charge/discharge curves of the AC//CuS ASC are shown in Fig. 4c. At a current density of 2 mA cm$^{-2}$, the volumetric capacitance of the AC//CuS ASC reached 3.54 F cm$^{-3}$, which is much higher capacitance in comparison with other reported ASCs.[28-30] Fig. 4d shows the Ragone plots of the AC//CuS ASC, revealing the energy and power density calculated from the discharge curves. Notably, the AC//CuS ASC exhibited the higher volumetric energy density of 1.11 mWh cm$^{-3}$ and the enhanced volumetric power density of 0.36 W cm$^{-3}$ in comparison with those of commercial energy storage devices and other reported pseudo-capacitive SCs.[31-36] In addition, even after 2000 charge/discharge cycles at a current density of 10 mA cm$^{-2}$ (Fig. 4e), the AC//CuS ASC showed a good cyclability with the high capacitance retention of 91.0 % and Coulombic efficiency of 97.7 %.

**CONCLUSIONS**

In summary, we demonstrated the CuS NW arrays directly grown on the Cu mesh successfully prepared via a facile liquid-solid chemical oxidation process and an anion exchange reaction for the enhanced pseudo-capacitive behavior properties in SCs. The obtained CuS NWs exhibited a superior areal capacitance of 378.0 mF cm$^{-2}$ at a current density of 2 mA cm$^{-2}$ and good rate performance as well as long-term electrochemical and structural stabilities during the galvanostatic charge/discharge process, compared to other reported Cu-based SCs. The improved electrochemical pseudo-capacitive behavior properties of the CuS NWs can be attributed to unique electrical and structural properties of the CuS NWs grown directly on Cu mesh current collector, providing synergistic benefits such as high electrical conductivity, good contact resistance, densely well-arrayed 1-D nanostructures with nanosized grains and high GBs, and the stable and favorite structure of a CuS phase. Furthermore, the AC//CuS ASC with a wide potential window of 1.5 V exhibited the enhanced volumetric energy density of 1.11 mWh cm$^{-3}$, the good cyclability of 91% up to 2000 cycles, and the high Coulombic efficiency of 97.7 %. Thus, it is expected that the CuS NWs directly grown on a Cu mesh current collector will be extensively utilized as highly stable and efficient pseudo-capacitive electrodes in SCs.

## EXPERIMENTAL

*Fabrication of 1-D Cu(OH)$_2$ and CuS NWs*

Copper hydroxide nanowires (Cu(OH)$_2$ NWs) were synthesized directly on the Cu mesh by using an aqueous solution of sodium hydroxide (NaOH, 97%) and ammonium persulfate ((NH$_4$)$_2$S$_2$O$_8$, 98%). The solution was prepared by mixing 20 mL of a 10 M NaOH solution, 10 mL of a 1 M (NH$_4$)$_2$S$_2$O$_8$ solution, and 22.5 mL of DI water. A pre-cleaned piece of a Cu mesh (1 x 1 cm$^2$ with thickness of 0.0267 cm, Alfa Aesar) was subsequently immersed into the mixed solution at room temperature. After the synthesis of Cu(OH)$_2$ NWs for 1h, the Cu mesh sample was taken out of the solution and then rinsed with DI water, followed by dried on a hot plate at 60 °C for 15 min. A light blue color of the mesh was observed, indicating that highly dense Cu(OH)$_2$ NWs were synthesized. In order to convert as-grown $Cu(OH)_2$ NWs to CuS NWs, a 0.2 M thiourea ($CH_4N_2S$, 99.0%) solution was dropped on the surface of $Cu(OH)_2$ NWs and then blown dry with nitrogen gas. After heating on the hot plate at 150 °C, the color of the sample was changed from a blue to dark brown color. It indicates that CuS NWs were completely transformed from $Cu(OH)_2$ NWs.

*Electrochemical Characterization*

The electrochemical properties of the as-prepared $Cu(OH)_2$ and CuS NWs directly grown on the Cu mesh were measured in a three-electrode system, consisting of as-prepared CuS NWs electrodes as a working electrode, Pt wire as a counter electrode, and Ag/AgCl (in saturated 3 M KCl) as a reference electrode, in order to analyze CV, galvanostatic charge/discharge, and EIS behavior using a potentiostat (PGSTAT302N, Metrohm, Autolab). For the fabrication of anode electrodes in the AC//CuS ASC, the slurry was prepared by mixing the active carbon as an active material, poly(vinylidene difluoride) as a binder, Ketjen black as a conductive material, and then was coated onto the compressed nickel foam as a current collector. All electrochemical results of the AC//CuS ASC were obtained using a two electrode system under 1.0 M KOH solution at room temperature.

**Acknowledgments**

This research was supported by the Industrial Fundamental Technology Development Program (10052745, Development of nano-sized (100nm) manganese ceramic material for high voltage pseudo-capacitor) funded by the Ministry of Trade, Industry and Energy (MOTIE) of Korea, and the European Research Council under the European Union's Seventh Framework Programme (FP/2007-2013) / Grant Agreement no. 685758, Project '1D-NEON'.

**Figure**

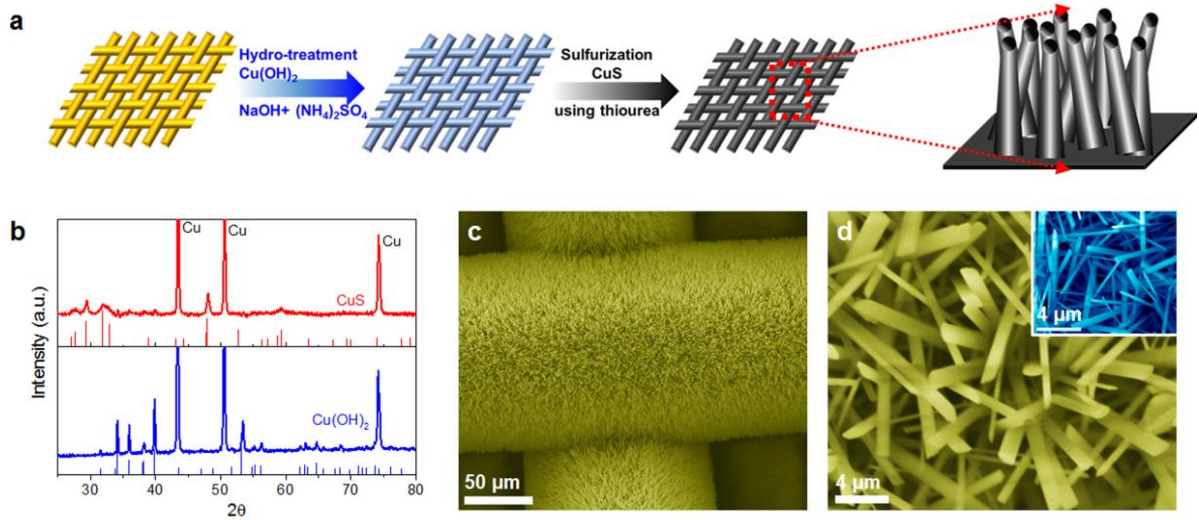

**Fig. 1.** (a) Schematic illustration of the synthesis process of CuS NWs. (b) XRD patterns of the as-prepared Cu(OH)$_2$ and CuS NWs. (c) A FE-SEM image of CuS NWs. (d) A high-magnification SEM image of CuS NWs. The inset indicates the high-magnification SEM image of Cu(OH)$_2$ NWs.

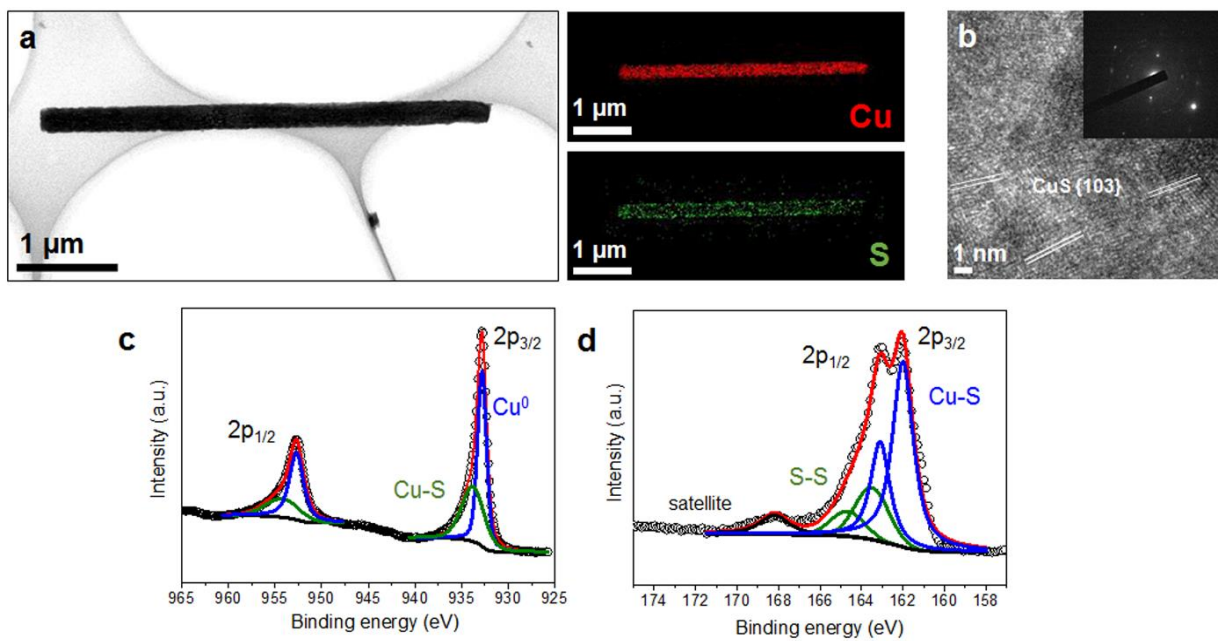

**Fig. 2.** (a) TEM (left) and corresponding EDX element mapping images (right) for the Cu (red) and S (green) in the CuS NW. (b) High-resolution TEM images of the CuS NW. The inset indicates the FFT pattern of a CuS NW. XPS (c) Cu 2p and (d) S 2p spectra of the CuS NWs.

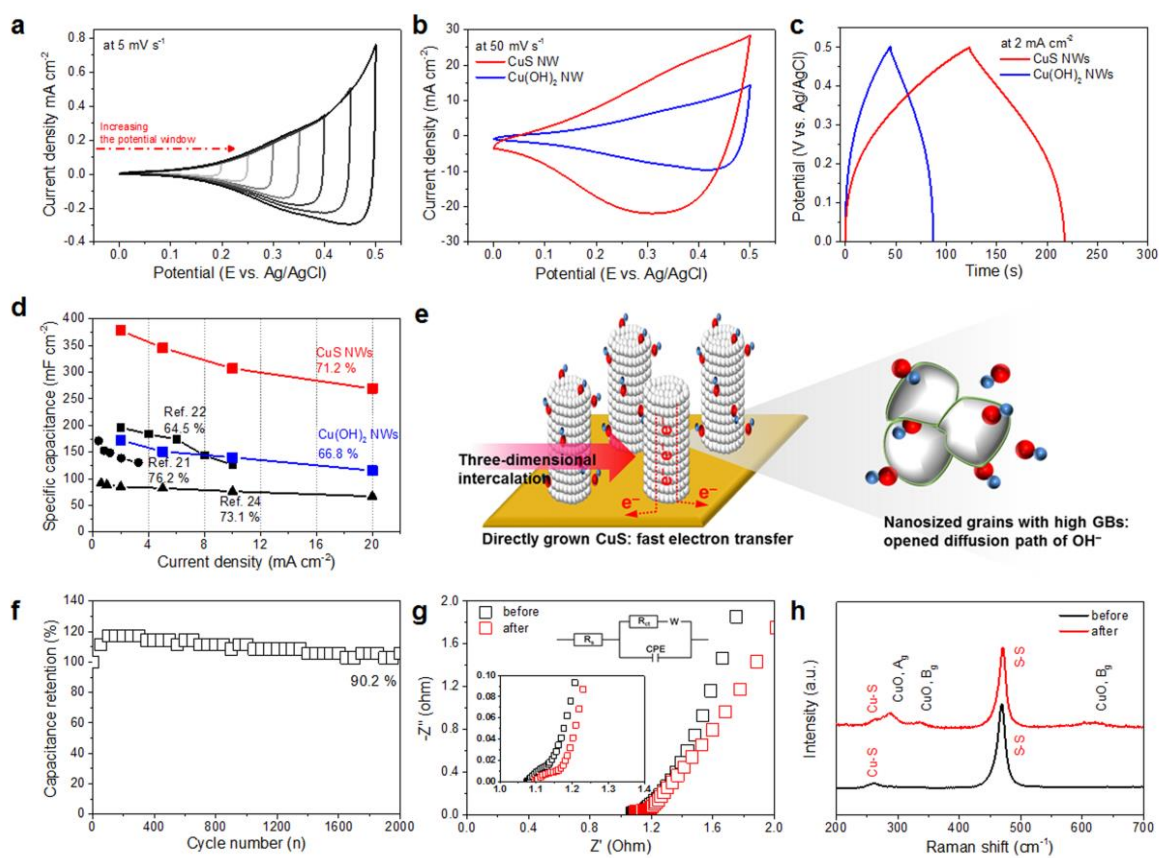

**Fig. 3.** (a) CVs of CuS NWs in different upper potential windows from 0.2 to 0.5 V. (b) CVs of CuS NWs and Cu(OH)$_2$ NWs at a scan rate of 50 mV s$^{-1}$. (c) Galvanostatic charge/discharge curves of CuS NWs and Cu(OH)$_2$ NWs at a current density of 2 mA cm$^{-1}$. (d) Specific capacitances of CuS NWs and Cu(OH)$_2$ NWs at different constant current densities in comparison with other reported Cu-based electrodes. (e) Schematic illustration related to pseudo-behavior properties of OH$^-$ ions in the CuS NW. (f) Cycle performance of CuS NWs at a current density of 10 mA cm$^{-2}$ up to 2000 cycling charge/discharge tests. (g) Nyquist plots and (h) Raman spectra of the CuS NWs before and after cycling tests. Insets in Fig. 3g indicate the expanded Nyquist plots and the equivalent circuit.

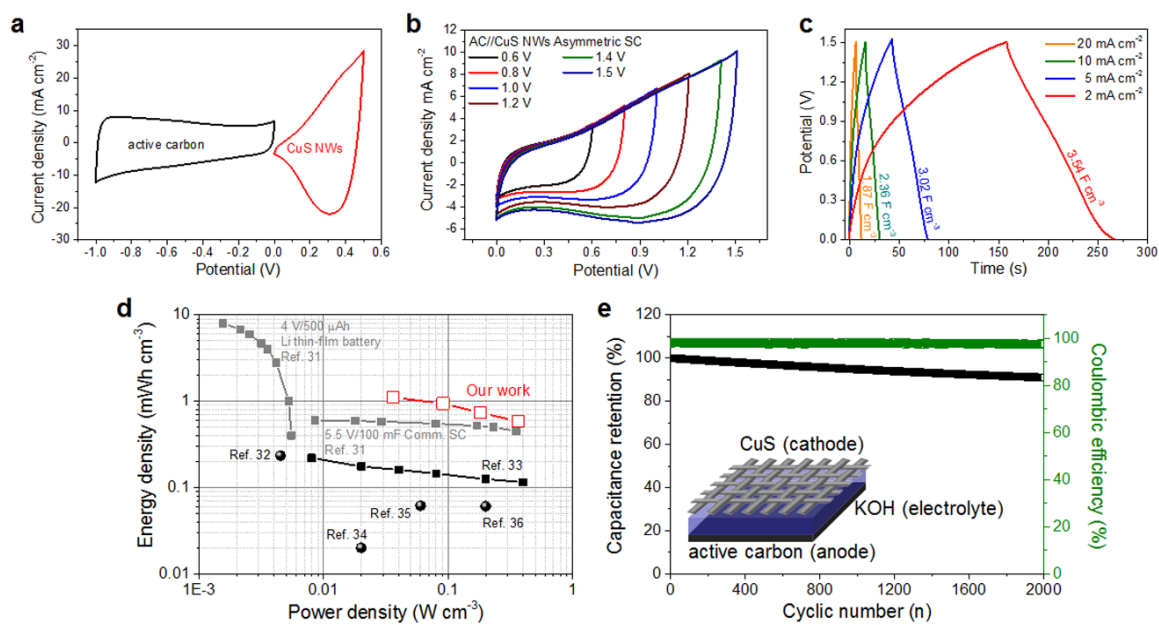

**Fig. 4.** (a) Comparative CVs of AC and CuS NWs at 50 mV s$^{-1}$. (b) CVs of the AC//CuS ASC in different upper potential windows from 0.6 to 1.5 V at a scan rate of 50 mV s$^{-1}$. (c) Galvanostatic charge/discharge curves of the AC//CuS ASC under various different current densities. (d) Ragone plots of the AC//CuS ASC and recently reported other pseudo-capacitive SCs. (e) Cycle performance of the AC//CuS ASC at a current density of 10 mA cm$^{-2}$ up to 2000 cycling charge/discharge tests.